\documentclass[twocolumn]{aastex701}

\usepackage{graphicx}
\usepackage{bm}  
\usepackage[colorlinks=true]{hyperref} 
\usepackage{gensymb}
\usepackage{array}
\usepackage{natbib}
\usepackage{amssymb, amsmath}
\begin{document}
\title{A Super-Eddington, Lensing-Magnified Quasar at $z=5.07$ observed with JWST}

\correspondingauthor{Katherine Panebianco}

\author[0009-0006-2158-9784]{Katherine Panebianco}
\email{kpanebia@mit.edu}
\affiliation{Department of Physics, Massachusetts Institute of Technology, 77 Massachusetts Ave., Cambridge, MA 02139, USA}

\author[0000-0002-5367-8021]{Minghao Yue}
\email{yuemh@arizona.edu}
\affiliation{Steward Observatory, University of Arizona, 933 N. Cherry Ave., Tucson, AZ 85718, USA}
\affiliation{MIT Kavli Institute for Astrophysics and Space Research, 77 Massachusetts Ave., Cambridge, MA 02139, USA}

\author[0000-0003-2895-6218]{Anna-Christina Eilers}
\email{eilers@mit.edu}
\affiliation{MIT Kavli Institute for Astrophysics and Space Research, 77 Massachusetts Ave., Cambridge, MA 02139, USA}
\affiliation{Department of Physics, Massachusetts Institute of Technology, 77 Massachusetts Ave., Cambridge, MA 02139, USA}

\author[0000-0003-3310-0131]{Xiaohui Fan}
\email{xfan@arizona.edu}
\affiliation{Steward Observatory, University of Arizona, 933 N. Cherry Ave., Tucson, AZ 85718, USA}

\author[0000-0002-7633-431X]{Feige Wang}
\email{fgwang@umich.edu}
\affiliation{Department of Astronomy, University of Michigan, 1085 S. University Ave., Ann Arbor, MI 48109, USA}

\author[0000-0001-5287-4242]{
Jinyi Yang}
\email{jyyangas@umich.edu}
\affiliation{Department of Astronomy, University of Michigan, 1085 S. University Ave., Ann Arbor, MI 48109, USA}

\author[0000-0003-3769-9559]{Robert A. Simcoe}
\email{rsimcoe@mit.edu}
\affiliation{MIT Kavli Institute for Astrophysics and Space Research, 77 Massachusetts Ave., Cambridge, MA 02139, USA}

%%%%%%%%%%%%%%%%%%%%%%%%%%%%%%%%%%%%

\begin{abstract}
    We present JWST/NIRCam F070W and F480M imaging for a quasar at $z = 5.07$, J0025--0145, which is magnified by a foreground lensing galaxy. 
    Existing Hubble Space Telescope {\em (HST)} imaging does not have sufficient spatial resolution to determine whether the background quasar is multiply imaged. 
    Exploiting the sharp PSF of the F070W band, we confirm that the background quasar can be well-described by a single point spread function (PSF), essentially ruling out the existence of multiple lensed images. We do not detect the quasar host galaxy in either the F070W or the F480M band.
    Using the {\em HST} and {\em JWST} photometry, we fit the Spectral Energy Distribution (SED) of the foreground galaxy. The estimated mass ($\log(M_{*} / M_{\odot}) = 11.15 \pm 0.16$) and redshift ($z_{\text{phot}} = 3.62_{-0.04}^{+0.06}$) of the foreground galaxy are consistent with a single-image lensing model.
    We estimate the maximum possible magnification of the quasar to be $\mu_{\text{max}} = 3.2$, which implies that the intrinsic Eddington ratio of the quasar is at least $\lambda_{\text{Edd}}^{\text{intrinsic}} > 4.9$. Therefore, J0025--0145 has one of the highest Eddington ratios among $z>5$ supermassive black holes known so far, suggesting the viability of super-Eddington growth for supermassive black holes in the early universe. 
\end{abstract}

%%%%%%%%%%%%%%%%%%%%%%%%%%%%%%%%%%%%
\section{Introduction}\label{sec:intro}

High-redshift quasars provide a unique opportunity to probe the properties of supermassive black holes (SMBHs) and their host galaxies \citep[for a recent review, see][]{fan_quasars_2023}.
The existence of such large SMBHs at high redshifts challenges current models of SMBH formation and growth in the early universe \citep[e.g.,][]{yang_probing_2021,eilers_eiger_2023}, placing lower bounds on SMBH seed masses and accretion rates \citep[e.g.,][]{mortlock_luminous_2011,banados_800-million-solar-mass_2018}. 
One possible explanation for the formation of these massive SMBHs in the early universe is super-Eddington accretion (\citealt{banados_800-million-solar-mass_2018}; \citealt{suh_super-eddington-accreting_2025}).
Therefore, finding super-Eddington accreting SMBHs will provide crucial insights into early SMBH growth.

To date, hundreds of quasars have been identified at $z > 5$ (e.g. \citealt{shen_gemini_2019}; \citealt{fan_quasars_2023}; \citealt{2016ApJS..227...11B}) and used to study early universe SMBHs (e.g. \citealt{yang_probing_2021}). 
A significant fraction of $z>5$ quasars has been predicted to be gravitationally lensed \citep[e.g.,][]{pacucci19}, and extensive efforts have been performed to find these high-redshift lensed quasars \citep[e.g.,][]{yue_survey_2023,andika23}.
Nevertheless, only one strongly lensed quasar has been confirmed at $z > 5$ (J0439+1634 at $z=6.52$, \citealt{fan_discovery_2019}). 
J0439+1634 provides an example of how gravitational lensing aides the study of high-redshift quasars; lensing magnification has enabled the measurements of faint signals and small-scale structures that are usually impossible to resolve for high-redshift quasars (\citealt{yang_far-infrared_2019}; \citealt{yue_alma_2021}; \citealt{yang_deep_2022}). 

Recently, \citet{yue_survey_2023} reported an intermediately lensed quasar at redshift $z=5.07$, J0025--0145. 
J0025--0145 was first reported as a luminous quasar with $M_{1450}=-28.63$ \citep{wang_survey_2016}.
Follow-up {\em Hubble Space Telescope (HST)} imaging reveals a foreground galaxy $\sim0\farcs6$ away from the quasar \citet{yue_survey_2023}. The existence of a foreground galaxy with such a close separation suggests that the background quasar must be magnified by gravitational lensing. When fitted as a single point source, the quasar shows a significant positive residual in the \textit{HST} image, implying that J0025--0145 might exhibit multiple lensed images \citep{yue_survey_2023}. However, \textit{HST} images do not have sufficient spatial resolution to confirm the existence of multiple lensed images and determine the lensing structure of J0025--0145. 

Interestingly, J0025--0145 has a high apparent Eddington ratio for the accretion of the central SMBH of $\lambda^{\text{apparent}}_{\text{Edd}} = 9$, which is one of the highest Eddington ratios for $z>5$ SMBHs ever reported. However, lensing magnification biases the apparent Eddington ratio as $\lambda^{\text{apparent}}_{\text{Edd}}=\lambda^{\text{intrinsic}}_{\text{Edd}}\times\mu^{0.5}$ \citep[see discussions in, e.g.,][]{fan_discovery_2019,yue_survey_2023}. As the lensing model of J0025--0145 remains undetermined, we cannot evaluate the intrinsic Eddington ratio of J0025--0145.

To fully determine the lensing structure of J0025--0145 and measure the magnification of the background quasar, we obtain {\em James Webb Space Telescope (JWST)} images in the F070W and F480M bands. The F070W image has a spatial resolution of $\sim0\farcs03$, about three times better than the {\em HST} image, which provides a much more stringent test for the multi-image scenario. Meanwhile, the F480M image covers the rest-frame wavelength $\lambda\sim8000\text{\AA}$ at the quasar's wavelength,  optimal for characterizing the quasar host galaxy emission.

The layout of the paper is as follows. 
In Section \ref{sec:data}, we provide the details of the \textit{JWST} observations and data reduction process. 
In Section \ref{sec:psf}, we model the point spread function (PSF) in both \textit{JWST} NIRCam filters and perform PSF subtraction to resolve the lensing structure of J0025--0145. 
In Section \ref{sec:discussion}, we discuss the lensing properties of J0025--0145 and use the \textit{JWST} observations and previous \textit{HST} observations to perform SED fitting on the foreground lensing galaxy. 
Further, we use the lensing properties for J0025-0145 to put a constraint on its magnification and Eddington ratio. 
We summarize all results in Section \ref{sec:summary}.
Throughout this paper, we assume a flat $\Lambda$CDM cosmology with $\Omega_M=0.3$ and $H_0=70\text{ km s}^{-1}\text{kpc}^{-1}$, and use AB magnitudes \citep{ABsystem}.

%%%%%%%%%%%%%%%%%%%%%%%%%%%%%%%%%%%%
\section{Data}\label{sec:data}

\begin{figure}%[!ht]
    \centering
    \includegraphics[width=1\linewidth]{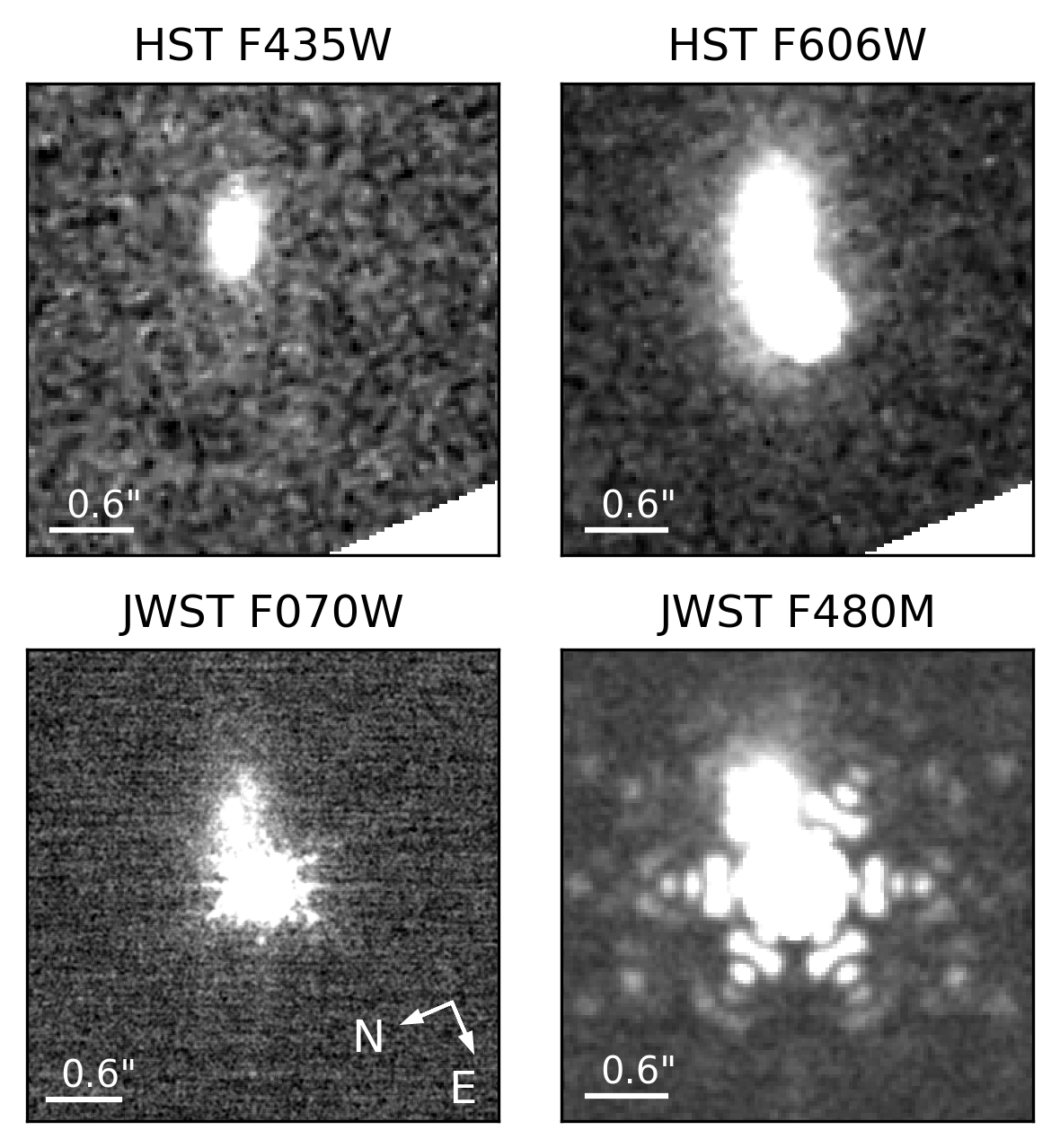}
    \caption{Images of J0025--0145. 
    Top, from left to right: \textit{HST} ACS/WFC F435W and \textit{HST} ACS/WFC F606W.
    Bottom, from left to right: \textit{JWST} NIRCam F070W and \textit{JWST} NIRCam F480M.
    Both the quasar and the foreground galaxy are visible in the \textit{HST} F606W, \textit{JWST} F070W, and \textit{JWST} F480M filters. 
    Only the foreground galaxy is visible in the \textit{HST} F435W filter.}
    \label{fig:data}
\end{figure}

\subsection{HST Imaging}

J0025--0145 was observed by {\em HST} Advanced Camera for Surveys Wide-Field Camera (ACS/WFC) in the F435W band and the F606W band (Program \#16460, PI: Yue). We refer the readers to \citet{yue_survey_2023} for the data reduction and analysis of these observations. The top panels of Figure \ref{fig:data} show the {\em HST} images of J0025--0145. The quasar is clearly detected in the F606W image but has no flux in the F435W image, due to the absorption of the intergalactic medium (IGM). Meanwhile, the galaxy appears in both the F435W and the F606W filter. This feature confirms that the galaxy is in the foreground, and the background quasar must be magnified by the galaxy.

\citet{yue_survey_2023} reported a positive residual flux when fitting the quasar as a point source. The residual flux can be explained by multiple lensed images unresolved in the {\em HST} image or significant errors of the PSF model.

\subsection{JWST Imaging} \label{sec:jwst_imaging}

J0025--0145 was observed by {\em JWST} NIRCam in the F070W and the F480M bands (Program \#3017, PI: Yue). To improve the sampling of the sharp F070W PSF, we adopt a 4-point INTRAMODULEBOX primary dither pattern and a 9-point STANDARD subpixel dither pattern. The total exposure time is $1932.624 \, \text{s}$ for both filters.
We reduce the data using {\texttt{jwst}} version 1.17.1, and adapt the $1/f$ noise removal and background subtraction algorithm from the CEERS NIRCam pipeline\footnote{https://github.com/ceers/ceers-nircam}. 

The bottom panels of Figure \ref{fig:data} show the \textit{JWST} images of J0025--0145. The pixel scales are $0\farcs01$ in the F070W band and $0\farcs03$ in the F480M band. At first glance, the quasar appears to be a single point source in both bands. Given the sharp PSFs of NIRCam ($\sim0\farcs03$ in the F070W band), these images already disfavor the multi-image scenario. We will further test whether J0025--0145 contains multiple PSF components in Section \ref{sec:psf}.

%%%%%%%%%%%%%%%%%%%%%%%%%%%%%%%%%%%%
\section{Fitting the Images of J0025--0145}\label{sec:psf}

To determine whether the background quasar in J0025--0145 is multiply-imaged, we first build PSF models for the F070W and F480M images, then fit the images to test how many PSF components are needed to describe the images.

\begin{figure*}[!ht]
    \centering
    \includegraphics[width=0.95\linewidth]{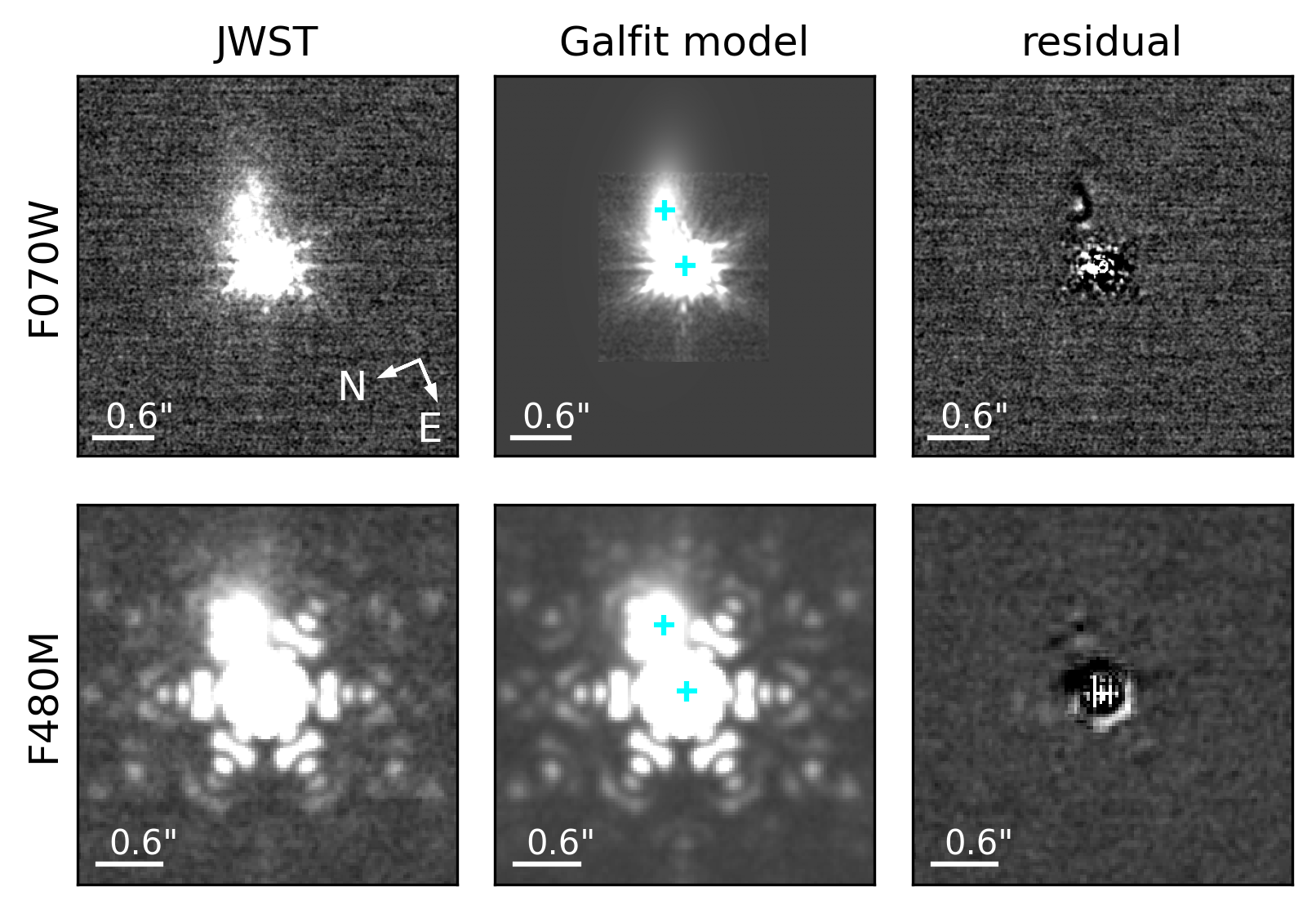}
    \caption{Best \textit{Galfit} models and residuals for \textit{JWST} NIRCam F070W (top) and F480M (bottom) filters. 
    The left column shows a cutout of the quasar and lensing galaxy, the middle column shows the \textit{Galfit} model, with the quasar modeled as a PSF and the lensing galaxy modeled with a S\'ersic profile, and the right column shows the residuals after subtracting the model.
    The quasar and galaxy locations are labeled in the model with a cross.
    For both filters, a single PSF image fits the quasar best.}
    \label{fig:galfit}
\end{figure*}

\setlength{\tabcolsep}{15pt}
\renewcommand{\arraystretch}{1.2}
\begin{deluxetable*}{c c c}
    \centering
    \caption{\textit{Galfit} best fit parameters}
    \label{tab:galfit}
    \startdata
    \vspace{-0.35cm}
    \\
     & F070W & F480M \\
    \hline
    PSF Image & $(\Delta x, \Delta y)\tablenotemark{1} \equiv (0, 0)$ & $(0,0)$ \\ 
    & $ m _{\text{F070W}} = 18.7000 \pm 0.0005$ & $ m_{\text{F480M}} = 17.5784 \pm 0.0003$ \\
    \hline 
    Lensing Galaxy & $(-0\farcs2201, 0\farcs5883)$ & $(-0\farcs2127, 0\farcs6273)$\\
    & $m _{\text{F070W}} = 21.06 \pm 0.02$ & $ m_{\text{F480M}} = 20.17 \pm 0.02$ \\
    \hline
    S\'ersic Parameters & $R_e = 0\farcs481 \pm 0\farcs008$ & $R_e = 0\farcs301 \pm 0\farcs008$ \\
    & $n = 1.16 \pm 0.02$ & $n = 4.315 \pm 0.095$ \\
    & $q = 0.437 \pm 0.004$ & $q = 0.530 \pm 0.005$ \\
    & $\textrm{PA} = -6.77 \pm 0.45 \degree$ & $\textrm{PA} = -1.45 \pm 0.36 \degree$ \\
    \enddata
    \tablenotetext{1}{J0025--0145 has (RA, Dec) = (00:25:26.83, --01:45:32.48) \citep{yue_survey_2023}.}
\end{deluxetable*}

\subsection{PSF Modeling}

We model the PSF in both \textit{JWST} NIRCam filters using bright stars from publicly available archival NIRCam images. 
We note that the image taken in our program does not have any bright stars suitable for PSF modeling. 
Thus, we download NIRCam images from the MAST archive that are publicly available by June 2025, with exposure times longer than 1,000 seconds. We only include images that cover the NRCB1 aperture (for the F070W band) or the NRCB5 aperture (for the F480M band), where J0025--0145 is located. We reduce these images using the same method as described in Section \ref{sec:jwst_imaging}.  This step yields 117 images in the F070W band and 104 images in the F480M band.

We then run \texttt{SExtractor} \citep{1996A&AS..117..393B} to extract sources in these images. 
We select objects with $0\farcs030<$FLUX\_RADIUS$<0\farcs034$ (for the F070W band) and $0\farcs093<$FLUX\_RADIUS$<0\farcs099$ (for the F480M band) as potential PSF stars. 
Since the PSF of infrared detectors might be flux-dependent \citep[e.g.,  the brighter-fatter effect; e.g.,][]{Goudfrooij24}, we restrict our PSF stars to $18.2<m_\text{F070W}<19.2$ and $17<m_\text{F480M}<18$, matching the magnitudes of the quasar (Section \ref{sec:image_fitting}). 
Finally, we visually inspect all the selected stars to exclude any with bright companions or those severely affected by bad pixels. This process yields 38 PSF stars in the F070W band and 14 in the F480M band.
 
We then use \texttt{PSFEx} \citep{bertin_automated_2011} to construct the PSF models. \texttt{PSFEx} models the spatial variation of the PSF across the detector, and has been shown to provide reliable PSFs for NIRCam imaging \citep{zhuang_characterization_2023}. 
Specifically,  we feed the PSF star lists described above to \texttt{PSFEx}, and fit the PSF model as a linear function of the $x-$ and $y-$position on the detector. We then use \texttt{ShapePipe} (\citealt{guinot_shapepipe_2022}; \citealt{farrens_shapepipe_2022}) to interpolate the PSF at the position of J0025--0145, and use these PSFs for image fitting. 

The PSF stars and the final PSF models will be available online upon the acceptance of this paper.

\subsection{Image Fitting} \label{sec:image_fitting}

We use \textit{Galfit} \citep{peng_detailed_2002} to fit the images of J0025--0145.
\textit{Galfit} is a software that uses analytical profiles to model 2-dimensional images.
We first fit the data in the F070W filter to resolve the number of lensed quasar images that are present, taking advantage of its better spatial resolution. 
We model lensed quasar images as point sources, model the lensing galaxy as a S\'ersic profile, and include a sky background component.  
Figure \ref{fig:galfit} shows the best-fit result for the single PSF model. The quasar is well-described by a single PSF, and we find no signs of multiple lensed images. 
We also see no signs of the host galaxy in the residual image.

We also attempt to 
fit the quasar as two PSFs in the F070W filter. 
The best-fit model has one PSF at approximately the same magnitude as the best-fit single PSF model ($\approx 18.7$), with the other PSF magnitude much dimmer ($\approx 22.8$). The fainter PSF is $\approx 0\farcs08$ away from the brighter one and is located along the horizontal PSF spike. This suggests that the second PSF component is produced by PSF model errors instead of a second lensed image.
We also note that
the probability of producing two lensed images at such a small separation and high flux ratio is extremely low ($\lesssim0.5\%$ among all strong lensing systems, according to the model in \citealt{yue22}). Therefore, we conclude that the background quasar in J0025--0145 is not multiply-imaged, and that 
the positive flux residual seen in the {\em HST} image is a result of the inaccurate {\em HST} PSF model.

After confirming that there is only one lensed image, we fit the data in the F480M filter to determine whether the extended host galaxy is visible.
We model the data with a single PSF image for the quasar and a S\'ersic profile for the lensing galaxy. 
The bottom panel of Figure \ref{fig:galfit} shows the result; 
we do not see any signs of an extended host galaxy in the residual image.

We present the parameters from the \textit{Galfit} best fit model in Table \ref{tab:galfit}.
We note that the S\'ersic parameters for the lensing galaxy are quite different in the F070W and F480M filters, indicating different galaxy morphology in these wavelength ranges, possibly due to a differential dust attenuation for the lensing galaxy.

%%%%%%%%%%%%%%%%%%%%%%%%%%%%%%%%%%%%
\section{Implications on the Lensing Properties of J0025--0145}\label{sec:discussion}

As the background quasar in J0025--0145 is not multiply-lensed, we do not have sufficient information to fit a lensing model to J0025--0145. 
Instead, we infer the properties of this lensing system based on the position, photometry, and morphology of the foreground galaxy.

\subsection{The Lensing Galaxy}

The lensing galaxy has NIRCam $m_\text{F070W}=21.06 \pm 0.02$ and $m_\text{F480M}=20.17 \pm 0.02$ (Table \ref{tab:galfit}). \citet{yue_survey_2023} measured the ACS/WFC F606W magnitude of the lensing galaxy to be $m_\text{F606W}=21.34 \pm 0.03$, but did not provide its F435W magnitude. 
We thus fit the ACS/WFC F435W image as a single S\'ersic profile using {\em Galfit}, which yields $m_\text{F435W}=23.28\pm0.03$ for the galaxy. Previous spectroscopic observations were not able to determine the redshift of the lensing galaxy \citep{yue_survey_2023}.
We thus estimate the mass and redshift of the lensing galaxy by SED fitting.

Specifically, we use \texttt{Prospector} \citep{prospector}. The input photometry includes {\em HST} ACS/WFC F435W and F606W bands, as well as {\em JWST} NIRCam F070W and F480M bands. 
The free parameters and their priors of this SED model include: 
(1) the redshift of the galaxy with a flat prior at $[0, 5.07]$;
(2) the stellar mass $M_*$ with a log-uniform prior at $[10^8M_\odot, 10^{12}M_\odot]$;
(3) the stellar metallicity $\log (Z/Z_\odot)$ with a uniform prior at $[-2, 0.2]$;
(4) the starting time of the star formation $t_\text{age}$ with a uniform prior at $[0, 10\,\text{Gyr}]$;
(5) the exponential decay timescale $\tau$ with a uniform prior at $[0.01\,\text{Myr}, 20\,\text{Myr}]$;
(6) the dust attenuation (quantified as the optical depth at 5500{\AA}, $\tau_{5500}$) with a uniform prior at $[0, 2]$; 
(7) the gas-phase metallicity $\log (Z_g/Z_\odot)$ with a uniform prior at $[-2, 0.5]$; 
and (8) the ionization parameter $\log U$ with a uniform prior at $[-3, 1]$. 
We assume a delayed-$\tau$ model for the star formation history (SFH), 
i.e., $\text{SFR}(t)\propto te^{-t/\tau}$.
We use a Chabrier initial mass function \citep{chabrier03} and assume a dust attenuation following the \citet{calzetti03} law.

\begin{figure}
    \centering
    \includegraphics[width=1\linewidth]{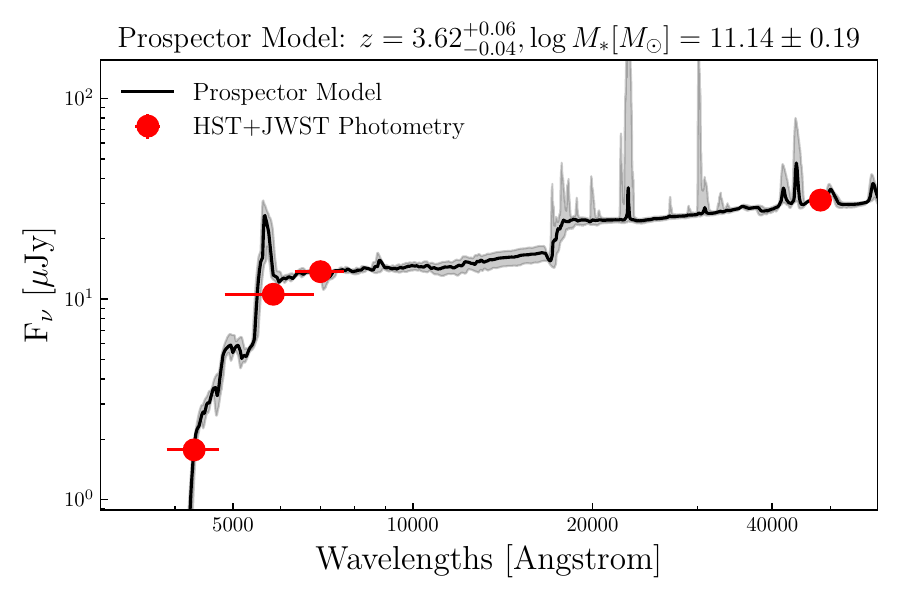}
    \caption{The \texttt{Prospector} model for the foreground lensing galaxy. The red dots are the {\em HST} and {\em JWST} photometric points. The black line gives the median of the posterior model spectrum, and the gray shaded area represents the 16th and 84th percentiles.} 
    \label{fig:lens_sed_fitting}
\end{figure}

Figure \ref{fig:lens_sed_fitting} demonstrates the result of SED fitting. The lensing galaxy has a photometric redshift of $z_\text{phot}=3.62^{+0.06}_{-0.04}$, and a stellar mass of $\log (M_*/M_\odot)=11.15\pm0.16$. We note that \texttt{Prospector} does not find a low-redshift solution for the SED model.

With the redshift and mass of the galaxy, we can evaluate the Einstein radius $(\theta_\text{E})$ of this galaxy. We follow the method in \citet{yue22}; specifically, we first evaluate the dynamical mass of the galaxy as $M_\text{dyn}=M_*/0.557$ \citep[e.g.,][]{bezanson11,bezanson12}, then convert $M_\text{dyn}$ to the velocity dispersion using the relation by applying the virial theorem:
\begin{equation} \label{eq:virial}
    \sigma = \sqrt{\frac{ GM_\text{dyn}}{K R_e}}
\end{equation}
where the scaling factor $K$ depends on the morphology of the galaxy and has typical values of $K\sim5-8$ \citep{bertin02}. We adopt the typical scaling factor $K=6$ for early-type galaxies \citep{cappellari06,nn16}. 
Following these relations and using the SED fitting posterior, we get $\sigma=285_{-53}^{+68}\text{km s}^{-1}$ for the lensing galaxy.

We then compute the Einstein radius as
\begin{equation} \label{eq:thetaE}
    \theta_E=4\pi\left(\frac{\sigma}{c}\right)^2\frac{D_\text{ds}}{D_\text{s}}
\end{equation}
where 
$D_\text{s}$ is the angular diameter distance from the observer to the source, 
and $D_\text{ds}$ is the angular diameter distance from the deflector to the source.
This analysis yields $\theta_\text{E}=0\farcs27^{+0\farcs10}_{-0\farcs08}$. As we will soon show in Section \ref{sec:lensmodel}, the estimated Einstein radius is consistent with our finding that the quasar is not multiply imaged.

We end this Section by commenting on the photometric redshift of the deflector galaxy. 
Among all the known lensing systems, the most distant deflector galaxy is at $z_d\approx2$ \citep{vd24}. Therefore, J0025--0145 might have the most distant deflector galaxy among all known lensing systems. For a background source at $z=5.07$, a deflector at $z\approx3.62$ is rare but still possible \citep[e.g.,][]{yue22b}. If the deflector redshift is confirmed by follow-up spectroscopy, J0025--0145 will provide a rare opportunity to investigate the mass profiles (via lens modeling) and the circumgalactic medium properties (via absorption lines in the quasar spectrum) of a high-redshift deflector galaxy \citep[e.g.,][]{shajib21,huyan25}.

\subsection{The Lensing Structure}\label{sec:lensmodel}

Since J0025--0145 shows only one image of the background quasar, 
it is impossible to pin the Einstein radius and the magnification by lens modeling. The reason is that the observed position and flux of the quasar can be produced by a range of Einstein radii, corresponding to different source positions and fluxes (and thus magnifications).
Here, we investigate the lensing properties, especially the magnification, as a function of the galaxy's Einstein radius.

We use a Singular Isothermal Ellipse (SIE) model to describe the mass profile of the lensing galaxy, and assume that the SIE has the same position, ellipticity, and position angle as the F480M light profile.
We consider Einstein radii of $0<\theta_\text{E}<0\farcs4$; for each value of $\theta_E$, we find the source-plane quasar position that reproduces the observed image-plane position. We use {\textsc{Glafic}} \citep{glafic_software} to perform lens modeling.

The result suggests that the observed image-plane position can be produced by any $\theta_\text{E}<0\farcs38$, and the magnification $\mu$ increases with $\theta_\text{E}$. Thus, the maximum possible magnification without multiple lensed images is $\mu_\text{max}=3.2$. Lensing models with $\theta_\text{E}>0\farcs38$ will produce at least two images, inconsistent with the finding of NIRCam image fitting. This $\theta_\text{E}$ range agrees with the finding from SED fitting.

We also note that, for lensing models that produce two lensed images, the separation between the two images is at least $0\farcs4$. As described in Section \ref{sec:image_fitting}, no PSF components are seen at such large separations.

\subsection{A Quasar with Super-Eddington Accretion}

The Eddington ratio of a quasar scales as $\lambda_{\text{Edd}} \propto L_{\text{bol}} / M_{\text{BH}}$, where $L_{\text{bol}}$ is the bolometric luminosity of the quasar and $M_{\text{BH}}$ is the mass of the SMBH.
The SMBH mass $M_{\text{BH}}$ is typically measured using broad emission line widths and luminosities \citep[e.g.,][]{vestergaard_mass_2009},
\begin{equation}
   \log M_{\text{BH}} = 0.5\log(\lambda L_{\lambda}) + 2\log(\text{FWHM})+ C
\end{equation}
where FWHM is the full-width half maximum of broad emission lines, and $C$ is a constant depending on the specific broad line.
Lensing magnification changes the apparent luminosities and has no impact on line widths. Consequently, the apparent Eddington ratio scales as $\lambda_{\text{Edd}}^\text{apparent} = \lambda_{\text{Edd}}^{\text{intrinsic}} \times \mu^{0.5}$ (e.g., \citealt{fan_discovery_2019}, \citealt{fujimoto_truth_2020}).  

Without correcting for lensing magnification, J0025--0145 has a high apparent Eddington ratio of $\lambda_{\text{Edd}} \approx 9$ \citep{yue_survey_2023}. 
Therefore, the maximum possible magnification $\mu_{\text{max}} = 3.2$ indicates an intrinsic Eddington ratio of at least $\lambda_{\text{Edd}}^{\text{intrinsic}} > 4.9$. 
This lower limit confirms that J0025--0145 has one of the highest Eddington ratios among $z>5$ SMBHs reported so far (e.g., \citealt{wu_catalog_2022}).

J0025--0145 suggests that super-Eddington growth with $\lambda_\text{Edd}\gtrsim5$ is a viable pathway for SMBH growth in the early Universe, especially for type-I luminous quasars. This super-Eddington growth channel allows SMBHs to gain mass in a shorter timescale compared to Eddington-limited accretion, which might contribute to the overmassive black holes identified in high-redshift quasars \citep[e.g.,][]{yue24,stone24,marshall25} and the ``young quasar" population with exceptionally short lifetimes \citep[e.g.,][]{eilers20,eilers21,eilers25,yue23young}.

\null
\null
%%%%%%%%%%%%%%%%%%%%%%%%%%%%%%%%%%%%
\section{Summary}\label{sec:summary}

In this paper, we study the lensing structure of J0025--0145, a gravitationally lensed quasar at redshift $z = 5.07$ with a foreground galaxy visible $0\farcs60$ away \citep{yue_survey_2023}. 
We present and fit observations of J0025--0145 in the \textit{JWST} NIRCam F070W and F480M filters, modeling the quasar image as a PSF and the nearby foreground galaxy as a S\'ersic profile. 
The main findings of our paper are summarized as follows:

\begin{enumerate}
    \item A single PSF fits J0025--0145 best in both the \textit{JWST} NIRCam F070W and F480M filters. We find no evidence of multiple lensed images in either filter, concluding that J0025--0145 is likely singly lensed. We see no extended arc of the quasar's host galaxy in either filter. 
    \item Using the \textit{JWST} observations and previous \textit{HST} observations of J0025--0145 and its lensing galaxy, we perform SED fitting and estimate properties of the lensing galaxy. We find that the lensing galaxy has photometric redshift $z_{\text{phot}} = 3.62_{-0.04}^{+0.06}$ and stellar mass $\log(M_{*} / M_{\odot}) = 11.15 \pm 0.16$. We compute the Einstein radius of this galaxy as $\theta_\text{E}=0\farcs27^{+0\farcs10}_{-0\farcs08}$.
    \item We find that the J0025--0145 has a maximium possible magnification of $\mu_{\text{max}} = 3.2$. This maximum possible magnification places a lower bound on the intrinsic Eddington ratio of $\lambda_{\text{Edd}}^{\text{intrinsic}} > 4.9$ and confirms J0025--0145 as one of the most extreme accreting SMBHs at $z > 5$. 
    SMBHs at high-redshift that are accreting with such a high Eddington ratio are extremely interesting to study in order to understand the rapid growth of massive black holes in the very early universe.
\end{enumerate}

J0025--0145 is among the most extreme super-Eddington accreting SMBHs at $z>5$. 
Thanks to lensing magnification, J0025--0145 provides one of the best opportunities to characterize a super-Eddington accreting SMBH in the early Universe.

%%%%%%%%%%%%%%%%%%%%%%%%%%%%%%%%%%%%

\begin{acknowledgments}
This work is based on observations made with the NASA/ESA/CSA James Webb Space Telescope. The data were obtained from the Mikulski Archive for Space Telescopes at the Space Telescope Science Institute, which is operated by the Association of Universities for Research in Astronomy, Inc., under NASA contract NAS 5-03127 for JWST. These observations are associated with program ID $\#3017$.

\end{acknowledgments}

\bibliography{references}{}
\bibliographystyle{aasjournalv7}

\end{document}